\documentclass{LT23auth}
\usepackage{graphicx}

\begin{document}

\begin{frontmatter}

\title{Freezing of stripes in lightly-doped La$_{2-x}$Sr$_x$CuO$_4$ as
manifested in magnetic and transport properties of untwinned single
crystals}

\author[address1]{A. N. Lavrov \thanksref{thank1}},
\author[address1]{Seiki Komiya},
\author[address1]{Yoichi Ando}

\address[address1]{Central Research Institute of Electric Power Industry,
2-11-1 Iwado-kita, Komae, Tokyo 201-8511, Japan}

\thanks[thank1]{Corresponding author. E-mail: lavrov@criepi.denken.or.jp}

\begin{abstract}
Resistivity and magnetization measurements are used for studying the {\it
transverse} sliding of AF domain boundaries in lightly doped
La$_{2-x}$Sr$_x$CuO$_4$. We discuss that it is the freezing of the
transverse boundary motion that is responsible for the appearance of
``spin-glass'' features at low temperatures.
\end{abstract}

\end{frontmatter}

In high-$T_c$ cuprates, charges and spins in the CuO$_2$ planes tend to
self-organize in a peculiar striped manner, where the doped holes form
quasi-1D ``charged stripes'' separating antiferromagnetic (AF) domains
\cite{1D,theory,suscept,mobility,anis}. Manifestations of the
unidirectional AF domain (stripe) structure in lightly doped
La$_{2-x}$Sr$_x$CuO$_4$ (LSCO) have been found in neutron scattering
\cite{1D} and in such macroscopic properties as magnetic susceptibility
\cite{suscept} and resistivity \cite{mobility,anis}; in particular,
remarkable in-plane resistivity anisotropy \cite{anis} has shown that the
charge motion is actually facilitated along the stripe direction. Here we
report that the resistivity and magnetization can also be used for
studying the {\it transverse} sliding of the AF domain boundaries, and
show that the stripe freezing in lightly doped LSCO coincides with the
transition into the ``spin-glass'' state.

The details of LSCO crystal growth and detwinning (the crystals were
detwinned to avoid the stripe pinning by crystallographic twin boundaries)
along with details of measurements are described in Refs.
\cite{suscept,mobility,anis}.

\begin{figure}[!t]
\includegraphics*[width=1.0\linewidth]{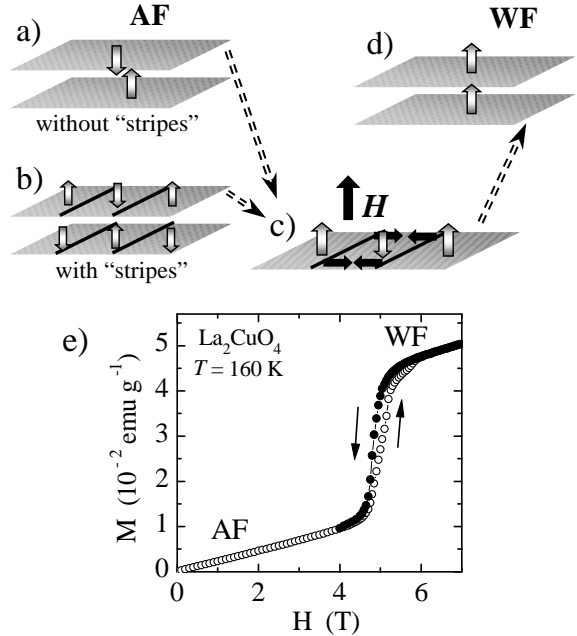}
\caption{(a-d) Motion of AF domain boundaries in LSCO upon the
weak-ferromagnetic transition; gray arrows indicate the direction of
canted moments, which is uniquely linked with the {\it phase} of the AF
order. (e) Magnetization behavior ($H\parallel c$) of La$_2$CuO$_4$
illustrating the transition into the WF state.}
\label{Fig.1}
\end{figure}

Owing to the spin canting induced by the Dzyaloshin\-skii-Moriya
interaction, the AF order in LSCO is always accompanied with a weak {\it
ferromagnetic} component \cite{MR_WF}. At zero magnetic field, the weak
ferromagnetism is hidden: the direction of canted moments depends on the
local {\it phase} of the AF order, and is opposite in neighboring CuO$_2$
planes (Fig. 1a). Apparently, if CuO$_2$ planes themselves contain AF
domains with opposite phases, the canted moments form a similar pattern in
the in-plane direction, changing their sign upon crossing the antiphase
AF-domain boundaries (Fig. 1b). A magnetic field applied along the
$c$-axis couples with the canted moments and eventually causes a
transition into the weak-ferromagnetic (WF) state \cite{MR_WF}, where the
phase of the AF order is unified and all the canted moments are aligned
along the field direction (Fig. 1d); the corresponding step-like increase
in the magnetization is illustrated in Fig. 1e. Whatever the initial
magnetic state is -- a homogeneous AF order (Fig. 1a) or a striped domain
structure (Fig. 1b) -- the transition into the WF state should involve the
transverse motion of the antiphase domain boundaries as shown in Fig. 1c.
This opens an intriguing possibility of studying the kinetics of the
transverse domain-boundary (stripe) sliding.

\begin{figure}[!t]
\leftskip5pt
\includegraphics*[width=0.90\linewidth]{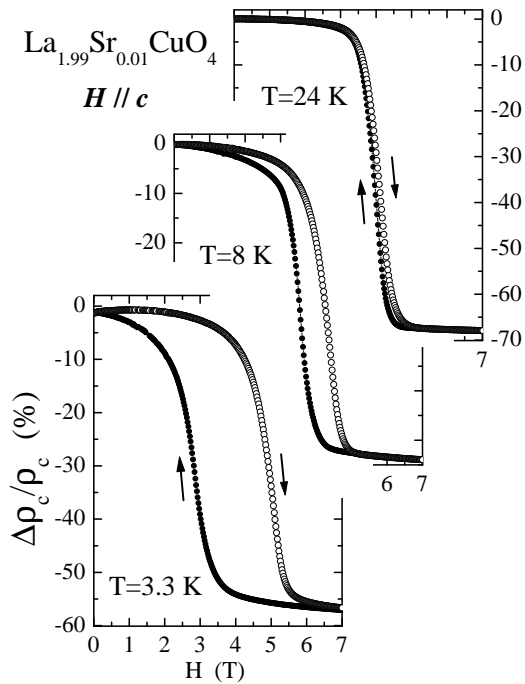}
\caption{The $c$-axis MR measured in a La$_{1.99}$Sr$_{0.01}$CuO$_4$
crystal upon increasing ($\circ$) and decreasing ($\bullet$) magnetic
field at a rate of 0.4 T/min.}
\label{Fig.2}
\end{figure}

A large change in the resistivity at the WF transition \cite{MR_WF} points
to the magnetoresistance (MR) as the most convenient probe to watch the
transition kinetics. Figure 2 illustrates the MR behavior in a
La$_{1.99}$Sr$_{0.01}$CuO$_4$ crystal upon increasing and decreasing the
magnetic field. While the WF transition is virtually reversible at high
temperatures, a significant hysteresis develops upon cooling, as the
pinning of the AF domain boundaries gains strength (Fig. 2). Using the
transition fields obtained from the MR data, we have sketched the magnetic
phase diagram (Fig. 3) which clearly shows that the transition region
between the AF and WF states broadens abruptly upon cooling below $\approx
14\,$K -- exactly the temperature where spin-glass features appear in
weak-field magnetization data \cite{suscept}. Apparently, at low
temperatures the mobility of the AF domain boundaries becomes inhibited so
that the equilibrium state cannot be reached over the experimental time
scale.

\begin{figure}[!t]
\begin{center}\leavevmode
\includegraphics*[width=0.83\linewidth]{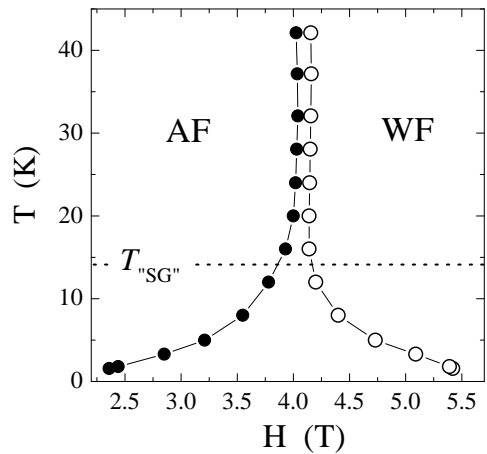}
\caption{Magnetic phase diagram of La$_{1.99}$Sr$_{0.01}$CuO$_4$
obtained from the MR data in Fig.2. The dashed line indicates the
``spin-glass'' transition temperature.}
\label{Fig.3}
\end{center}
\end{figure}

The following physical picture can be drawn from the obtained results: At
high temperatures, the AF domain boundaries are mobile and fluctuating in
the transverse direction; however, at low temperatures their pinning
quickly gains strength, bringing about the freezing of the AF domain
structure, and corresponding memory features in transport and magnetic
properties.

\vspace{12pt}
{\bf Acknowledgements}
\vspace{12pt}

We thank K. Segawa for invaluable technical assistance.

\end{document}